\documentclass[apjl,iop,twocolappendix]{emulateapj}
\usepackage[colorlinks,citecolor=blue,linkcolor=blue,urlcolor=blue,dvips]{hyperref}

\newcommand {\eg} {{\it e.g.}}

\newcommand {\be} {\begin{equation}}
\newcommand {\ee} {\end{equation}}
\newcommand {\bea} {\begin{eqnarray}}
\newcommand {\eea} {\end{eqnarray}}

\begin{document}

\title{Reconciling models of luminous blazars with magnetic fluxes 
determined by radio core shift measurements}

\shorttitle{Blazar models with the `core-shift'  magnetic fluxes}

\author{Krzysztof~Nalewajko\altaffilmark{1,2,3}, Marek Sikora\altaffilmark{4,5}, and Mitchell C. Begelman\altaffilmark{1,6}}

\shortauthors{Nalewajko, Sikora \& Begelman}

\altaffiltext{1}{JILA, University of Colorado and National Institute of Standards and Technology, 440 UCB, Boulder, CO 80309, USA}
\altaffiltext{2}{Kavli Institute for Particle Astrophysics and Cosmology, SLAC National Accelerator Laboratory, Stanford University, 2575 Sand Hill Road M/S 29, Menlo Park, CA 94025, USA; {\tt knalew@stanford.edu}}
\altaffiltext{3}{NASA Einstein Postdoctoral Fellow}
\altaffiltext{4}{Nicolaus Copernicus Astronomical Center, Bartycka 18, 00-716
Warsaw, Poland}
\altaffiltext{5}{JILA Visiting Fellow}
\altaffiltext{6}{Department of Astrophysical and Planetary Sciences, University of Colorado, 391 UCB, Boulder, CO 80309, USA}

\begin{abstract}
Estimates of magnetic field strength in relativistic jets of active galactic nuclei (AGN), obtained by measuring the frequency-dependent radio core location, imply that the total magnetic fluxes in those jets are consistent with the predictions of the magnetically-arrested disk (MAD) scenario of jet formation.
On the other hand, the magnetic field strength determines the luminosity of the synchrotron radiation, which forms the low-energy bump of the observed blazar spectral energy distribution (SED).
The SEDs of the most powerful blazars are strongly dominated by the high-energy bump, which is most likely due to the external radiation Compton (ERC) mechanism.
This high Compton dominance may be difficult to reconcile with the MAD scenario, unless 1) the geometry of external radiation sources (broad-line region, hot-dust torus) is quasi-spherical rather than flat, or 2) most gamma-ray radiation is produced in jet regions of low magnetization, e.g., in magnetic reconnection layers or in fast jet spines.
\end{abstract}


\section{Introduction}
\label{sec_intro}

Relativistic jets of active galactic nuclei (AGN) are launched by spinning black holes (BH) or accretion disks \citep{1977MNRAS.179..433B,1982MNRAS.199..883B}.
In radio galaxies and radio-loud quasars they are relativistic and reach powers comparable to
or sometimes even exceeding the accretion powers \citep{1991Natur.349..138R,2011MNRAS.411..901G,2011ApJ...728L..17P}.
Launching such powerful jets requires magnetic fluxes  which cannot be
developed by dynamo mechanisms in standard, radiation-dominated accretion disks \citep{1997MNRAS.292..887G}.
However, such fluxes are expected in the magnetically-arrested disk (MAD) scenario \citep[and references therein]{2003PASJ...55L..69N,2012MNRAS.423.3083M}.
In this case jets are produced most likely by the Blandford-Znajek mechanism, and the required strong 
net magnetic flux is expected to be accumulated on the BH  by   the advection 
of magnetic fields from external regions \citep[and references therein]{2013ApJ...764L..24S}.
Starting as Poynting flux-dominated outflows, these jets are smoothly accelerated as they convert energy from magnetic to kinetic and their 
magnetization $\sigma = B'^2/(4\pi w)$, where $w$ is the relativistic enthalpy density, drops with distance \citep[and references therein]{2007MNRAS.380...51K,2009ApJ...699.1789T}.
The jet acceleration becomes inefficient when $\sigma \lesssim 1$, from which point on, in the ideal magnetohydrodynamical (MHD) picture, $\sigma$ decreases logarithmically with distance \citep{2010MNRAS.402..353L}.
This transition between the two acceleration regimes may happen already at the distance of $10^3\;R_{\rm g} \sim 0.05\;{\rm pc}$ (here $R_{\rm g} = GM_{\rm bh}/c^2$ is the gravitational radius of the black hole of mass $M_{\rm bh} \sim 10^9M_\odot$) \citep{2007MNRAS.380...51K}.

AGN jets produce large amounts of non-thermal radiation, which is relativistically enhanced by $2-3$ orders of magnitude in blazars, where the jets are oriented close to the line of sight \citep{1979ApJ...232...34B}. This radiation is generally thought to be produced at a distance scale $0.01\;{\rm pc} < r < 10\;{\rm pc}$, but in many cases it can be narrowed to $0.1\;{\rm pc} < r < 1\;{\rm pc}$ (see \citealt{2014ApJ...789..161N} for a recent discussion). This main emission and dissipation region is referred to as the blazar zone. Modeling of the spectral energy distributions (SED) of blazars can be used to constrain the jet composition in the blazar zone \citep[\eg,][]{2000ApJ...534..109S}. The most luminous blazars, belonging to the class of flat-spectrum radio quasars (FSRQs), are strongly dominated by the gamma-ray emission, which is thought to be produced by the external-radiation-Comptonization (ERC) mechanism \citep{1992A&A...256L..27D,1994ApJ...421..153S,2009ApJ...704...38S}, the efficiency of which depends on the energy density of external radiation fields, mainly the broad emission lines (BEL) and the thermal radiation of the dusty torus. The very high apparent gamma-ray luminosities of FSRQs, at times exceeding $L_\gamma \sim 10^{48}\;{\rm erg\,s^{-1}}$ \citep{2011ApJ...733L..26A}, call for very high total jet powers $L_{\rm j} \sim 10^{46-47}\;{\rm erg\,s^{-1}}$ and high radiative efficiencies $\eta = L_\gamma/(\Gamma_{\rm j}^2L_{\rm j}) \sim 0.1$ \citep{2012Sci...338.1445N}, where $\Gamma_{\rm j} \sim 20$ is the jet Lorentz factor. On the other hand, most of the infrared/optical emission of blazars is due to the synchrotron mechanism, the efficiency of which depends on the local magnetic field strength. Observations of blazars with high dominance of the ERC component over synchrotron emission place strong constraints on $\sigma$ within the blazar zone, which we discuss in Section \ref{sec_q}.

Recently, the magnetic field strength scaled to the distance of $1\;{\rm pc}$ was estimated for a large sample of blazars and radio galaxies by using the core-shift technique \citep{2012A&A...545A.113P}. In this technique, the position of the radio core, assumed to be a photosphere due to the synchrotron self-absorption process \citep{1979ApJ...232...34B}, is measured in relation to sharp optically thin jet features as a function of observing frequency \citep{1998A&A...330...79L}. These magnetic field values were used to estimate the magnetic fluxes of jets $\Phi_{\rm j}$, which were compared to the theoretical magnetic fluxes threading the black holes $\Phi_{\rm bh}$ as predicted by the MAD scenario \citep{2014Natur.510..126Z}. The close agreement between $\Phi_{\rm j}$ and $\Phi_{\rm bh}$ strongly supports the MAD scenario for production of powerful AGN jets.
In Section \ref{sec_cores}, we show that this is equivalent to a relation $L_{\rm B} \sim L_{\rm d}$ between the magnetic jet power and accretion disk luminosity.

We identified a possible tension between the magnetic field strengths estimated from core-shift measurements, and the magnetic field strengths estimated from modeling the emission of the most Compton-dominated FSRQs. The latter tend to be lower by factor $\sim 3$, therefore, in Section \ref{sec_gamma} we consider dissipation sites that involve lower than average local magnetic field strengths: 1) magnetic reconnection layers, and 2) weakly-magnetized jet spines. We also emphasize the importance of the geometric distribution of external radiation sources, in particular that a flat geometry of the broad-line region (BLR) and/or the dusty torus makes the problem much worse. Our main results are summarized in Section \ref{sec_disc}.

\section{High Compton dominance in FSRQs}
\label{sec_q}

The SEDs of FSRQs are strongly dominated by the high-energy component peaking in the 10-100 MeV range, which is most naturally explained by the ERC model \citep{2009ApJ...704...38S}.
We define the Compton dominance parameter $q = L_{\rm ERC} / L_{\rm syn}$, where $L_{\rm ERC}$ and $L_{\rm syn}$ are the apparent luminosities of the ERC and synchrotron components, respectively, at their spectral peaks.
Numerous observations indicate that quite often $q \gtrsim 10$ for the brightest blazars \citep{2010ApJ...716...30A,2012A&A...537A..32A,2012A&A...541A.160G}.

On the other hand, if the ERC and synchrotron components are produced by the same population of electrons,\footnote{Statistically, blazars show significant correlation between the gamma-ray and optical fluxes \citep[\eg,][]{2014arXiv1404.5967C}. However, there are cases of poor correlation \citep[\eg,][]{2013ApJ...763L..11C}, in which one needs to consider multiple emitting regions. In such cases, one can focus on the main gamma-ray emitting region and place upper limits on the co-spatial synchrotron emission.} we can write $q \simeq u_{\rm ext}' / u_{\rm B}'$, where $u_{\rm ext}'$ and $u_{\rm B}'$ are the energy densities of the external radiation field and the magnetic field in the jet co-moving frame, respectively. The external radiation density can be parametrized as $u_{\rm ext}' = \zeta\Gamma^2L_{\rm d}/(4\pi cr^2)$, where $\zeta$ is a dimensionless parameter representing the details of reprocessing and beaming of the external radiation (see below), $\Gamma$ is the Lorentz factor of the emitting region, $L_{\rm d}$ is the accretion disk luminosity, and $r$ is the distance of the emitting region from the supermassive black hole. The magnetic energy density can be related to the jet magnetic power $L_{\rm B} = 2\pi R^2\Gamma^2u_{\rm B}'c$, where $R = \theta_{\rm j}r$ is the jet radius, and $\theta_{\rm j}$ is the half-opening angle of the jet. Gathering these relations together, we obtain the following constraint:
\be
\label{eq_q}
q = \left(\frac{\zeta}{0.005}\right)\left(\frac{\Gamma}{20}\right)^2(\Gamma\theta_{\rm j})^2\left(\frac{L_{\rm d}}{L_{\rm B}}\right)\,.
\ee
Written in such form, the above equation suggests the typical parameter values that we adopt as the starting point for further discussion.

The parameter $\zeta = \xi g_u$ includes the traditional covering factor $\xi$, and the geometric factor $g_u$ \citep{2013ApJ...779...68S}. The covering factor determines the total luminosity of the reprocessed accretion disk radiation, e.g., $L_{\rm BLR} = \xi_{\rm BLR}L_{\rm d}$. Typically, it is assumed that $\xi \simeq 0.1$, although there are many indications that it can be as high as $\xi \sim 0.4$ both for the BLR \citep{2007AJ....134.1061D,2009NewAR..53..140G}, and for the dusty tori \citep{2013MNRAS.429.1494R,2013ApJ...773...15W}.
The geometric factor depends on the geometric distribution of the reprocessing medium, and on the radial stratification of the covering factor.
As we show in the Appendix, even for a spherical distribution 
 $g_u < 0.7$, and for flattened distributions $g_u < 0.1$.
Recently, there has been increasing interest in flattened distributions of the BLR \citep{2012arXiv1209.2291T}, motivated mainly by observations of rapidly variable VHE emission from quasars \citep{2011ApJ...730L...8A}, and supported by direct observations \citep{2000ApJ...538L.103V,2011MNRAS.413...39D}.
The half-opening angle of the dusty tori is being estimated at $\sim 30^\circ$ \citep{2013ApJ...773...15W}.
Assuming that $g_u = \xi = 0.1$, we can expect that $\zeta$ can be as low as $0.01$. However, in the case of a quasi-spherical reprocessing medium with high covering factor, we may expect $\xi \simeq 0.4$ and $g_u \simeq 0.5$, and hence $\zeta \simeq 0.2$. High values of $q$ may thus require the presence of a dense, quasi-spherical medium reprocessing the central AGN radiation.

The Lorentz factors $\Gamma$ of blazar jets can be estimated from interferometric observations of apparent superluminal motions of radio features. Typical values for FSRQs are $10 < \Gamma < 40$ \citep{2009A&A...494..527H}. The jet collimation parameter $\Gamma\theta_{\rm j}$ should not exceed unity on both theoretical \citep{2009MNRAS.394.1182K} and observational \citep{2005AJ....130.1418J,2009A&A...507L..33P} grounds. Therefore, it is very unlikely that we could obtain $q > 10$ by increasing either the Lorentz factor or the collimation parameter.

Finally, the parameter $q$ can be increased by decreasing the magnetic jet power so that $L_{\rm B} < L_{\rm d}$.
If the jets are significantly magnetized, with $\sigma \simeq L_{\rm B}/(L_{\rm j}-L_{\rm B}) > 1$, we would expect that $L_{\rm B} \lesssim L_{\rm j}$, where $L_{\rm j}$ is the total jet power.
Observational evidence suggests that for the most powerful jets $L_{\rm j} \sim \dot{M}c^2 > L_{\rm d}$ (see Section \ref{sec_intro}).
This would be also consistent with the MAD scenario, in which it was demonstrated numerically that $L_{\rm j} \gtrsim \dot{M}c^2$ \citep{2011MNRAS.418L..79T}.
As we show in the next section, the requirement that $L_{\rm B} \sim L_{\rm d}$ is equivalent to the relation between the two magnetic fluxes $\Phi_{\rm j} \sim \Phi_{\rm bh}$ \citep{2014Natur.510..126Z}, therefore increasing $q$ by decreasing $L_{\rm B}$ \emph{globally} means a departure from the MAD scenario (in addition to departing from the core-shift measurements). However, one can still consider a \emph{local} decrease in the magnetic field strength in order to obtain a high $q$ (see Section \ref{sec_gamma}).

\section{Jet magnetic fields from core-shift measurements}
\label{sec_cores}

In this section we analyze the sample of blazars compiled by \cite{2014Natur.510..126Z}, for which magnetic field estimates $B_{\rm 1pc}'$ from core-shift measurements are available \citep{2012A&A...545A.113P}, as well as accretion disk luminosities $L_{\rm d}$, and black hole masses $M_{\rm bh}$.

\begin{figure}
\centering
\includegraphics[width=\columnwidth]{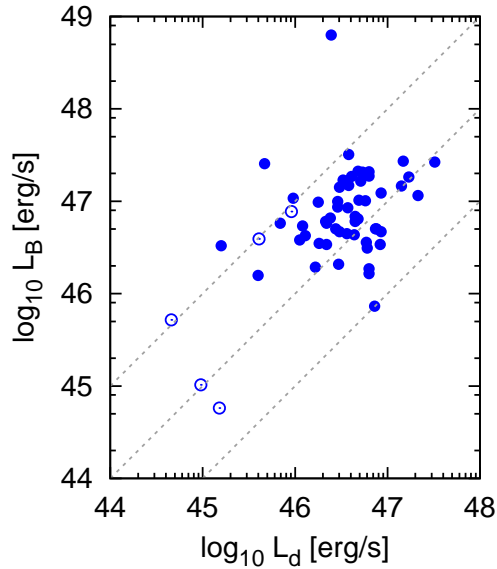}
\caption{Distribution of the accretion disk luminosity $L_{\rm d}$ vs. the magnetic jet power $L_{\rm B}$ for the sample of blazars (FSRQs -- \emph{solid points}; BL Lacs -- \emph{empty points}) compiled by \cite{2014Natur.510..126Z}. It is assumed that $\Gamma\theta_{\rm j} = 1$.}
\label{fig_Ld_LB}
\end{figure}

First, we estimate the magnetic jet power as $L_{\rm B} \simeq (c/4)(1\;{\rm pc})^2B_{\rm 1pc}'^2(\Gamma\theta_{\rm j})^2$.
In Figure \ref{fig_Ld_LB}, we show the distribution of $L_{\rm B}$ vs. $L_{\rm d}$ for the case of $\Gamma\theta_{\rm j} = 1$.
We note a substantial scatter in the $L_{\rm B}$ values, most of them falling in the range $0.2 < L_{\rm B}/L_{\rm d} < 20$.
The sources with $L_{\rm B} < L_{\rm d}$ may have $q > 1$, according to Eq. \ref{eq_q}.
However, very few sources in this sample can have $q > 10$ solely due to the low value of $L_{\rm B}/L_{\rm d}$.
As the magnetic jet power is a steep function of the jet collimation parameter $\Gamma\theta_{\rm j}$, allowing for $\Gamma\theta_{\rm j} < 1$ can substantially reduce $L_{\rm B}$.
However, since $q \propto (\Gamma\theta_{\rm j})^2/L_{\rm B}$ (Eq. \ref{eq_q}), the Compton dominance would not be affected by adopting a different value of $\Gamma\theta_{\rm j}$.

The correlation between $L_{\rm B}$ and $L_{\rm d}$ is much worse that the correlation between the two magnetic fluxes $\Phi_{\rm j}$ and $\Phi_{\rm bh}$ identified by \cite{2014Natur.510..126Z}.
Those magnetic fluxes can be written as:
\bea
\Phi_{\rm j} &\simeq& 8\pi(\Gamma\theta_{\rm j})f(a)R_{\rm g}B_{\rm 1pc}'(1\;{\rm pc}) \propto (\Gamma\theta_{\rm j})f(a)L_{\rm B}^{1/2}M_{\rm bh}\,,
\\
\Phi_{\rm bh} &\simeq& 50R_{\rm g}\left(\frac{L_{\rm d}}{\eta c}\right)^{1/2} \propto L_{\rm d}^{1/2}M_{\rm bh}\,,
\eea
where $\eta$ is the radiative efficiency of the accretion disk, $f(a)=[1+(1-a^2)^{1/2}]/a$, and $a$ is the dimensionless black hole spin. The very good correlation between the magnetic fluxes for $\Gamma\theta_{\rm j} = 1$, $a = 1$, and $\eta = 0.4$ can be partly explained by the fact that both fluxes are proportional to the black hole mass $M_{\rm bh}$. Because of the wide range of $M_{\rm bh}$ (about 3 orders of magnitude, $10^7 - 10^{10}M_\odot$), the relatively poor correlation between $L_{\rm B}$ and $L_{\rm d}$ is efficiently stretched along the lines of constant $L_{\rm B}/L_{\rm d}$. Also, since $\Phi_{\rm j}/\Phi_{\rm bh} \simeq (L_{\rm B}/L_{\rm d})^{1/2}$, the scatter between the $\Phi_{\rm j}/\Phi_{\rm bh}$ values is smaller than the scatter between the $L_{\rm B}/L_{\rm d}$ values.

\begin{figure}
\centering
\includegraphics[width=\columnwidth]{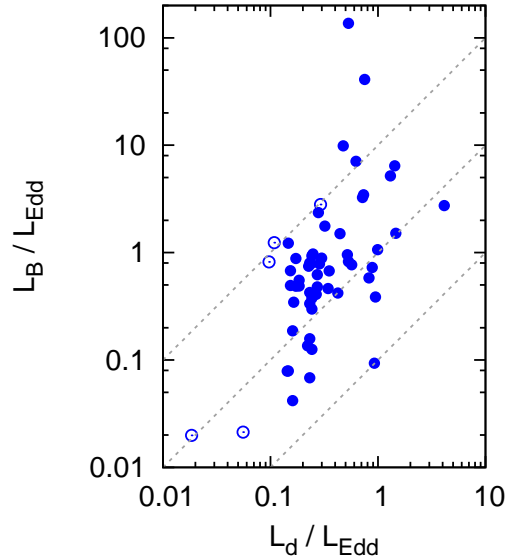}
\caption{Same as in Figure \ref{fig_Ld_LB}, but with both quantities scaled to the Eddington luminosity $L_{\rm Edd}$.}
\label{fig_LdEdd_LBEdd}
\end{figure}

In Figure \ref{fig_LdEdd_LBEdd}, we show the relation between $L_{\rm B}/L_{\rm Edd}$ and $L_{\rm d}/L_{\rm Edd}$, where $L_{\rm Edd} = 1.6\times 10^{38}(M_{\rm bh}/M_\odot)\;{\rm erg\,s^{-1}}$ is the Eddington luminosity. We note that the blazars in the sample compiled by \cite{2014Natur.510..126Z} occupy a narrow range of Eddington luminosity ratios, with $0.1 \lesssim L_{\rm d}/L_{\rm Edd} \lesssim 2$. All sources in the sample must have prominent broad emission lines in order to calculate both $L_{\rm d}$ and $M_{\rm bh}$. Because of this selection effect, effectively we have $L_{\rm d} \propto M_{\rm bh}$, $L_{\rm B} \propto M_{\rm bh}$, and the magnetic fluxes scale as $\Phi_{\rm j} \simeq \Phi_{\rm bh} \propto M_{\rm bh}^{3/2}$.

\section{Low-magnetization dissipation sites}
\label{sec_gamma}

We consider two potential mechanisms for obtaining reduced local magnetic field strengths in jets of typical magnetization $\sigma \sim 1$: one associated with reconnection layers and 
one related to the radial structure of magnetic fields across the jets.
Magnetic reconnection events are likely
to be triggered in  mildly relativistic turbulent plasma  
which is expected to be driven by current-driven instabilities \citep{1998ApJ...493..291B},
while stratification of the   
toroidal magnetic component across the jet may result from
balancing the magnetic stresses by the pressure of protons heated, e.g.,
by internal shocks.

\subsection{Reconnection layers}

Magnetic reconnection was proposed as an alternative dissipation mechanism for powering rapid high-amplitude gamma-ray flares in blazars \citep{2009MNRAS.395L..29G}. Efficient reconnection may reduce the local magnetic field strength by factor $\gtrsim 3$, which is necessary in order to achieve high Compton dominance $q > 10$, if the guide magnetic field component is more than 3 times smaller than the antiparallel magnetic field component.
Then, provided that magnetic energy released in the reconnection process is 
equally shared between protons and electrons \citep{2014AAinpress}, the electrons are injected with 
average random Lorentz factor $\bar\gamma_e \sim (m_p/m_e)\sigma \sim 10^3$. 
For $\sigma \sim 1$, these electrons  
can Comptonize external soft photons up to energy
$h\nu_{\rm ERC} \simeq (\Gamma/20)^2 (h\nu_{\rm ext}/10\,{\rm eV})$ GeV. In the case of ERC(BLR),
this is $\sim 100$ times larger than the energy of photons 
at typical $\gamma$-ray luminosity peaks, and in the case of ERC(IR) --- $\sim 3$ times larger.

In order to reconcile these energies with the peak location,
it is necessary to postulate an $e^+e^-$ pair content --- again assuming an equal energy partition between electrons and protons  --- $n_e/n_p \sim 100$ for ERC(BLR) and $n_e/n_p \sim 3$ for ERC(HDR),
where $n_e=n_{e^+} + n_{e^-}$. Noting that the efficiency of pair production at the characteristic distance scale of the BLR and beyond is very low
(the production of pairs
by absorption of the $\gamma$-rays by the UV photons requires an extension
of the $\gamma$-ray spectra above $\sim 30$ GeV),
such pairs must be produced at much lower distances,
close to the jet base,
where they can result from the absorption of the $\gamma$-rays by the X-rays produced in the accretion disk corona.
The significant pair content required in the reconnection scenario may explain why in the jet terminal shocks associated with 
radio-lobe hot spots, the observed low-energy break in the electron energy distribution is much lower than predicted 
by relativistic proton-electron shocks \citep[and references therein]{2007ApJ...662..213S,2013ApJ...767...12G}.

\subsection{Central core/spine}

If the jet has a lateral structure with a weakly magnetized core/spine \citep{2005A&A...432..401G} with $\sigma \sim 0.1$, and particles 
are accelerated by internal mildly relativistic shocks,
then the average energy gained by protons and
electrons (if shared equally) will be $\sim \eta_{\rm diss} m_p c^2/2$.
For a typical efficiency of energy dissipation in mildly relativistic shocks
$\eta_{\rm diss} \sim 0.1$ (see \citealt{2001MNRAS.325.1559S} for the internal shocks, and \citealt{2009MNRAS.392.1205N} for the reconfinement shocks)
this gives $\bar \gamma_e \sim 100$.
Such electrons boost external photons up to 
$\sim 40 (\Gamma/20)^2 (h\nu_{\rm ext}/10\,{\rm eV})$ MeV, which  
is roughly consistent with the location of the $\gamma$-ray spectral peaks.

\section{Conclusions}
\label{sec_disc}

Magnetic fluxes $\Phi_{\rm j}$ derived by measurements of radio core shifts in blazars \citep{2012A&A...545A.113P} are consistent with the maximum magnetic fluxes $\Phi_{\rm bh}$ predicted by the MAD model to thread the BH \citep{2014Natur.510..126Z}.
As we showed in Section \ref{sec_cores}, this is equivalent to the statement that magnetic jet power $L_{\rm B}$ is comparable to the accretion disk luminosity $L_{\rm d}$, which for total jet power $L_{\rm j} \sim L_{\rm d}$ implies typical jet magnetizations $\sigma \sim 1$.
However, as we showed in Section \ref{sec_q}, even in the case of a geometrically thick distribution of external radiation sources, significantly lower magnetization values
are required by radiation models of FSRQs in order to reproduce the high ratios $q$ of $\gamma$-ray to synchrotron luminosities.

This inconsistency can be resolved 
by noting that blazar jets need not be magnetically
homogeneous and uniform across the jet as commonly assumed.
As discussed in Section \ref{sec_gamma}, in realistic jet models there may exist regions with lower magnetization.
They can be generated by reconnection
driven in mildly relativistic turbulence.
They may also be associated
with jet cores filled with hot protons heated by internal shocks.
In both cases high values of $q$ are achievable,
and the energy of $\gamma$-ray
luminosity peaks can be reproduced --- in the shock scenario with a
proton-electron plasma, and in the reconnection scenario with significant pair content.

We also tentatively considered a possibility that the jet magnetic fields obtained from the radio core-shift measurements are overestimated. This could be the case if the radio cores are not photospheres due to the synchrotron self-absorption process, but rather they are due to a low-energy break in the electron distribution function. This idea will be developed in a future study.

\acknowledgments

We thank the reviewers and Andrzej Zdziarski for helpful comments on the manuscript.
M.S. thanks the JILA Fellows for their hospitality during the early stages of this project.
This project was partly supported by the NASA Fermi Guest Investigator program,
NASA Astrophysics Theory Program grant NNX14AB375,
and Polish NCN grant DEC-2011/01/B/ST9/04845.
K.N. was supported by NASA through Einstein Postdoctoral Fellowship grant number PF3-140130 awarded by the Chandra X-ray Center, which is operated by the Smithsonian Astrophysical Observatory for NASA under contract NAS8-03060.

\appendix

\begin{figure}[b]
\centering
\includegraphics[width=\columnwidth]{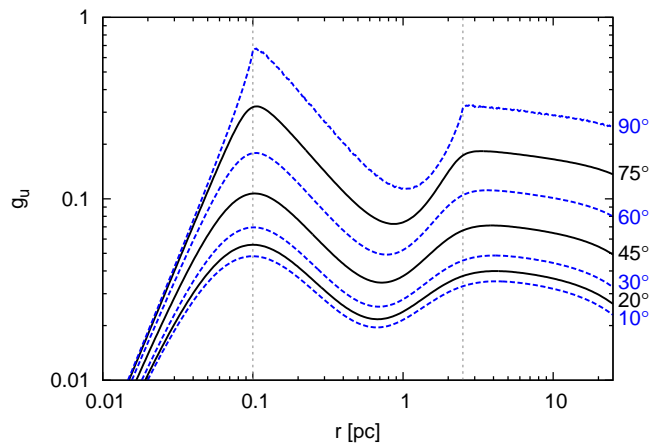}
\caption{The dependence of the geometrical correction factor $g_u$ for the external radiation density on the location $r$ along the jet and on the half-opening angle $\alpha_{\rm max}$ (its values for each curve are marked along the right edge) of the radiation source measured with respect to the accretion disk plane.}
\label{fig_gu}
\end{figure}

\section{Geometrical correction to the external radiation density}

In Section \ref{sec_q}, we parametrize the energy density of the external radiation fields, using a geometrical correction factor $g_u$, which was first introduced in \cite{2013ApJ...779...68S}. Given a particular geometrical model of the distribution of the medium producing the external radiation, we can integrate the external energy density $u_{\rm ext}'(r)$ along the jet in the co-moving frame, taking into account the exact distance and beaming factor for each volume element of the medium. Then, we calculate $g_u = 4\pi cr^2u_{\rm ext}'/(\xi\Gamma^2L_{\rm d})$.

We adopt here a specific geometry of the reprocessing medium (either BLR or the dusty torus), presented in Appendix A of \cite{2014ApJ...789..161N}. The optical depth gradient ${\rm d}\tau / {\rm d}r$ is assumed to scale like $r^{-2}$ for the BLR, and roughly like $r^{-1}$ for the dusty torus. We also adopt the covering factors $\xi_{\rm BLR} = \xi_{\rm IR} = 0.1$, and typical values for the inner radii $r_{\rm BLR}$ and $r_{\rm IR}$ of the BLR and the torus, respectively, from \cite{2009ApJ...704...38S}. The main variable is the half-opening angle $\alpha_{\rm max}$ of the medium measured from the accretion disk plane.

In Figure \ref{fig_gu}, we show the functions $g_u(r)$ for several values of $\alpha_{\rm max}$. We find that close to the characteristic radii $g_u \simeq 0.04$ for $\alpha_{\rm max} = 10^\circ$, $g_u \simeq 0.08$ for $\alpha_{\rm max} = 45^\circ$, and $g_u \simeq 0.2$ for $\alpha_{\rm max} = 75^\circ$. This indicates that widely different geometries of the reprocessing medium may change $g_u$, and thus $q$, by a factor $\sim 5$.


\begin{thebibliography}{}
      
\bibitem[Abdo et al.(2010)]{2010ApJ...716...30A}
Abdo, A.~A., Ackermann, M., Agudo, I., et al.\ 2010, \apj, 716, 30 

\bibitem[{Abdo} {et~al.}(2011)]{2011ApJ...733L..26A}
{Abdo}, A.~A., {Ackermann}, M., {Ajello}, M., {et~al.}, 2011, ApJ, 733, L26

\bibitem[{Aleksi{\'c}} {et~al.}(2011)]{2011ApJ...730L...8A}
{Aleksi{\'c}}, J., {et~al.}, 2011, ApJ, 730, L8

\bibitem[Arshakian et~al.(2012)]{2012A&A...537A..32A}
Arshakian, T.~G., Le{\'o}n-Tavares, J., B{\"o}ttcher, M., et al.\ 2012, A\&A, 537, A32

\bibitem[Begelman(1998)]{1998ApJ...493..291B}
Begelman, M.~C., 1998, ApJ, 493, 291

\bibitem[{Blandford} \& {K\"onigl}(1979)]{1979ApJ...232...34B}
{Blandford}, R.~D., {K\"onigl}, A., 1979, ApJ, 232, 34

\bibitem[{Blandford} \& {Payne}(1982)]{1982MNRAS.199..883B}
{Blandford}, R.~D., \& {Payne}, D.~G., 1982, MNRAS, 199, 883

\bibitem[{Blandford} \& {Znajek}(1977)]{1977MNRAS.179..433B}
{Blandford}, R.~D., \& {Znajek}, R.~L., 1977, MNRAS, 179, 433

\bibitem[Chatterjee et~al.(2013)]{2013ApJ...763L..11C}
Chatterjee, R., Fossati, G., Urry, C.~M., et~al.\ 2013, ApJ, 763, L11

\bibitem[Cohen et~al.(2014)]{2014arXiv1404.5967C}
Cohen, D.~P., Romani, R.~W., Filippenko, A.~V., et~al.\ 2014, arXiv:1404.5967

\bibitem[Decarli et al.(2011)]{2011MNRAS.413...39D}
Decarli, R., Dotti, M., \& Treves, A.\ 2011, \mnras, 413, 39 

\bibitem[{Dermer} {et~al.}(1992)]{1992A&A...256L..27D}
{Dermer}, C.~D., {Schlickeiser}, R., \& {Mastichiadis}, A., 1992, A\&A, 256, L27

\bibitem[Dunn et~al.(2007)]{2007AJ....134.1061D}
Dunn, J.~P., Crenshaw, D.~M., Kraemer, S.~B., \& Gabel, J.~R.\ 2007, AJ, 134, 1061

\bibitem[Gaskell(2009)]{2009NewAR..53..140G}
Gaskell, C.~M.\ 2009, NewAR, 53, 140

\bibitem[{Ghisellini} {et~al.}(2005)]{2005A&A...432..401G}
{Ghisellini}, G., {Tavecchio}, F., \& {Chiaberge}, M., 2005, A\&A, 432, 401

\bibitem[Ghisellini et~al.(2011)]{2011MNRAS.411..901G}
Ghisellini, G., Tagliaferri, G., Foschini, L., et al.\ 2011, \mnras, 411, 901 

\bibitem[Ghosh \& Abramowicz(1997)]{1997MNRAS.292..887G}
Ghosh, P., \& Abramowicz, M.~A.\ 1997, \mnras, 292, 887

\bibitem[Giannios et al.(2009)]{2009MNRAS.395L..29G}
Giannios, D., Uzdensky, D.~A., \& Begelman, M.~C.\ 2009, \mnras, 395, L29 

\bibitem[Giommi et~al.(2012)]{2012A&A...541A.160G}
Giommi, P., Polenta, G., L{\"a}hteenm{\"a}ki, A., et~al.\ 2012, \aap, 541, A160

\bibitem[Godfrey \& Shabala(2013)]{2013ApJ...767...12G}
Godfrey, L.~E.~H., \& Shabala, S.~S.\ 2013, \apj, 767, 12 

\bibitem[{Hovatta} {et~al.}(2009)]{2009A&A...494..527H}
{Hovatta}, T., {Valtaoja}, E., {Tornikoski}, M., {L{\"a}hteenm{\"a}ki}, A., 2009, A\&A, 494, 527

\bibitem[{Jorstad} {et~al.}(2005)]{2005AJ....130.1418J}
{Jorstad}, S.~G., {Marscher}, A.~P., {Lister}, M.~L., 2005, AJ, 130, 1418

\bibitem[{Komissarov} {et~al.}(2007)]{2007MNRAS.380...51K}
{Komissarov}, S.~S., {Barkov}, M.~V., {Vlahakis}, N., {K{\"o}nigl}, A., 2007, MNRAS, 380, 51

\bibitem[{Komissarov} {et~al.}(2009)]{2009MNRAS.394.1182K}
{Komissarov} S.~S., {Vlahakis} N., {K{\"o}nigl} A., {Barkov} M.~V., 2009, MNRAS, 394, 1182

\bibitem[Lobanov(1998)]{1998A&A...330...79L}
Lobanov, A.~P.\ 1998, \aap, 330, 79 

\bibitem[Lyubarsky(2010)]{2010MNRAS.402..353L}
Lyubarsky, Y.~E.\ 2010, \mnras, 402, 353 

\bibitem[McKinney et al.(2012)]{2012MNRAS.423.3083M}
McKinney, J.~C., Tchekhovskoy, A., \& Blandford, R.~D.\ 2012, \mnras, 423, 3083 

\bibitem[Melzani et~al.(2014)]{2014AAinpress}
Malzani, M., Walder, R., Folini, D. et~al. 2014, arXiv:1405.2938

\bibitem[{Nalewajko} \& {Sikora}(2009)]{2009MNRAS.392.1205N}
{Nalewajko}, K., {Sikora}, M., 2009, MNRAS, 392, 1205

\bibitem[Nalewajko et~al.(2014)]{2014ApJ...789..161N}
Nalewajko, K., Begelman, M.~C., \& Sikora, M.\ 2014, ApJ, 789, 161

\bibitem[Narayan et al.(2003)]{2003PASJ...55L..69N}
Narayan, R., Igumenshchev, I.~V., \& Abramowicz, M.~A.\ 2003, \pasj, 55, L69 

\bibitem[Nemmen et al.(2012)]{2012Sci...338.1445N} 
Nemmen, R.~S., Georganopoulos, M., Guiriec, S., et al.\ 2012, Science, 338, 1445

\bibitem[Punsly(2011)]{2011ApJ...728L..17P}
Punsly, B.\ 2011, \apjl, 728, L17 

\bibitem[{Pushkarev} {et~al.}(2009)]{2009A&A...507L..33P}
{Pushkarev}, A.~B., {Kovalev}, Y.~Y., {Lister}, M.~L., {Savolainen}, T., 2009, A\&A, 507, L33

\bibitem[Pushkarev et al.(2012)]{2012A&A...545A.113P}
Pushkarev, A.~B., Hovatta, T., Kovalev, Y.~Y., et al.\ 2012, \aap, 545, A113 

\bibitem[Rawlings \& Saunders(1991)]{1991Natur.349..138R}
Rawlings, S., \& Saunders, R.\ 1991, \nat, 349, 138 

\bibitem[Roseboom et~al.(2013)]{2013MNRAS.429.1494R}
Roseboom, I.~G., Lawrence, A., Elvis, M., et~al.\ 2013, MNRAS, 429, 1494

\bibitem[Sikora \& Begelman(2013)]{2013ApJ...764L..24S}
Sikora, M., \& Begelman, M.~C.\ 2013, \apjl, 764, L24

\bibitem[Sikora \& Madejski(2000)]{2000ApJ...534..109S}
Sikora, M., \& Madejski, G.\ 2000, ApJ, 534, 109

\bibitem[{Sikora} {et~al.}(1994)]{1994ApJ...421..153S}
{Sikora}, M., {Begelman}, M.~C., \& {Rees}, M.~J., 1994, ApJ, 421, 153

\bibitem[{Sikora} {et~al.}(2009)]{2009ApJ...704...38S}
{Sikora}, M., {Stawarz}, \L., {Moderski}, R., {Nalewajko}, K., \& {Madejski}, G.~M., 2009, ApJ, 704, 38

\bibitem[Sikora et~al.(2013)]{2013ApJ...779...68S}
Sikora, M., Janiak, M., Nalewajko, K., Madejski, G.~M., \& Moderski, R.\ 2013, ApJ, 779, 68
 
\bibitem[{Spada} {et~al.}(2001)]{2001MNRAS.325.1559S}
{Spada}, M., {Ghisellini}, G., {Lazzati}, D., \& {Celotti}, A., 2001, MNRAS, 325, 1559

\bibitem[Stawarz et al.(2007)]{2007ApJ...662..213S}
Stawarz, {\L}., Cheung, C.~C., Harris, D.~E., \& Ostrowski, M.\ 2007, \apj, 662, 213

\bibitem[Tavecchio \& Ghisellini(2012)]{2012arXiv1209.2291T}
Tavecchio, F., \& Ghisellini, G.\ 2012, arXiv:1209.2291

\bibitem[Tchekhovskoy et al.(2009)]{2009ApJ...699.1789T}
Tchekhovskoy, A., McKinney, J.~C., \& Narayan, R.\ 2009, \apj, 699, 1789 

\bibitem[{Tchekhovskoy} {et~al.}(2011)]{2011MNRAS.418L..79T}
{Tchekhovskoy}, A., {Narayan}, R., {McKinney}, J.~C., 2011, MNRAS, 418, L79

\bibitem[Vestergaard et al.(2000)]{2000ApJ...538L.103V}
Vestergaard, M., Wilkes, B.~J., \& Barthel, P.~D.\ 2000, \apjl, 538, L103

\bibitem[Wilkes et al.(2013)]{2013ApJ...773...15W}
Wilkes, B.~J., Kuraszkiewicz, J., Haas, M., et al.\ 2013, \apj, 773, 15 

\bibitem[Zamaninasab et~al.(2014)]{2014Natur.510..126Z}
Zamaninasab, M., Clausen-Brown, E., Savolainen, T., \& Tchekhovskoy, A.\ 2014, \nat, 510, 126

\end{thebibliography}
\end{document}